\documentclass[Physsubmission, Phys]{SciPost}

\hypersetup{
    colorlinks,
    linkcolor={red!50!black},
    citecolor={blue!50!black},
    urlcolor={blue!80!black}
}

\usepackage[export]{adjustbox}
\usepackage{subcaption}

\usepackage[bitstream-charter]{mathdesign}
\usepackage{xspace}
\usepackage{graphicx}        
\usepackage{rotating}
\usepackage{geometry} 
\usepackage{epsfig}

\DeclareSymbolFont{usualmathcal}{OMS}{cmsy}{m}{n}
\DeclareSymbolFontAlphabet{\mathcal}{usualmathcal}

\DeclareGraphicsExtensions{.pdf,.png,.jpg}

\newcommand{\app}{\emph{Astropart. Phys. }}

\newcommand{\etal}{et al.}
\newcommand{\nima}{\emph{Nucl. Instrum. Methods Phys. Res., Sect. A }}

\begin{document}

\begin{center}
\Large \textbf{
The full coverage approach to the detection of Extensive Air Showers}
\end{center}

\begin{center}
Giuseppe Di Sciascio\textsuperscript{$\star$}
\end{center}

\begin{center}
{\bf *} INFN - Roma Tor Vergata, Roma, Italy \\
E-mail: giuseppe.disciascio@roma2.infn.it
\end{center}

\begin{center}
\today
\end{center}


\definecolor{palegray}{gray}{0.95}
\begin{center}
\colorbox{palegray}{
  \begin{tabular}{rr}
  \begin{minipage}{0.1\textwidth}
    \includegraphics[width=30mm]{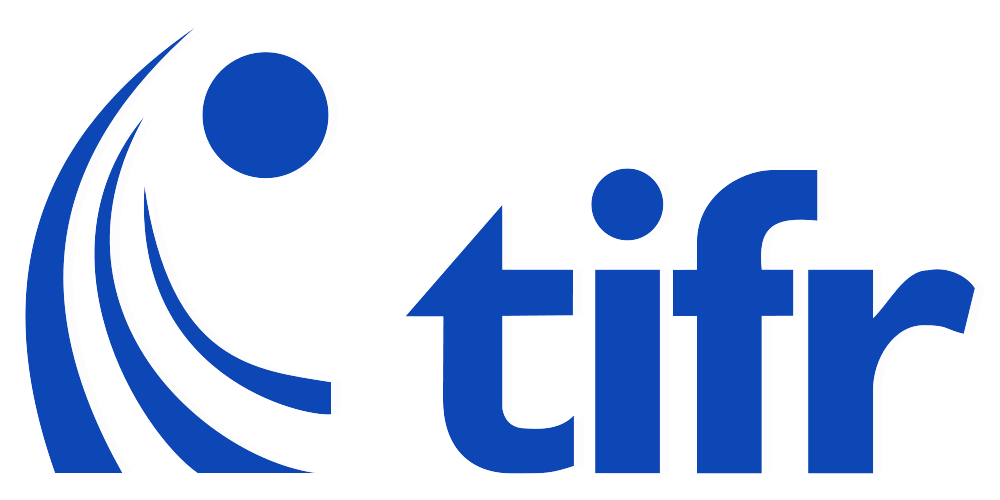}
  \end{minipage}
  &
  \begin{minipage}{0.85\textwidth}
    \begin{center}
    {\it 21st International Symposium on Very High Energy Cosmic Ray Interactions (ISVHECRI 2022)}\\
    {\it Online, 23-27 May 2022} \\
    \doi{10.21468/SciPostPhysProc.?}\\
    \end{center}
  \end{minipage}
\end{tabular}
}
\end{center}

\section*{Abstract}
{\bf
A shower array exploiting the full coverage approach with a high segmentation of the readout allow to image the front of atmospheric showers with unprecedented resolution and detail. The grid distance determines the energy threshold (small energy showers are lost in the gap between detectors) and the quality of the shower sampling.
Therefore, this experimental solution is needed to detect showers with a threshold in the 100 GeV range.
The full coverage approach has been exploited in the ARGO-YBJ experiment. In this contribution we will summarise the advantages of this technique and discuss possible applications in new wide field of view detectors.
}

\vspace{10pt}
\noindent\rule{\textwidth}{1pt}
\tableofcontents\thispagestyle{fancy}
\noindent\rule{\textwidth}{1pt}
\vspace{10pt}

\section{Introduction}
\label{sec:intro}

Gamma-ray astronomy above a few hundreds of GeV can be studied in a statistically significant way only with ground-based detectors exploiting the different components of Extensive Air Showers (EAS).
The classical approach is to detect, with suitable telescopes, the Cherenkov radiation produced in atmosphere by secondary charged particles in air showers.
Another possibility is to sample the charged particles at ground with arrays of detectors.
The main characteristics and differences between air shower arrays and Imaging Atmospheric Cherenkov Telescopes (IACT) are summarised in the Table \ref{tab:one}.

%
\begin{table}[h]
\vspace{0.5cm}
\centering
\caption{{\bf Characteristics of Air Shower Arrays and IACTs}}
\begin{tabular}{|c|c|c|}
\hline
\hline
  &  Air Shower Array & IACT  \\
\hline
 \hline
Energy Threshold  &  Low ($\approx$100 GeV) & Very low ($\approx$10 GeV)  \\
\hline
Max Energy  &  Very High ($\approx$10$^{15}$ eV) & Limited ($\approx$100 TeV)  \\
\hline
Field of View  &  Very Large ($\approx$2 sr) & Limited ($\approx$5 deg)  \\
\hline
Duty-cycle  &  Very large ($\approx$100\%) & Very Small ($\approx$15\%)  \\
\hline
Energy Resolution  &  Modest ($\approx$100 - 20\%) & Good ($\approx$15\%)  \\
\hline
Background rejection  &  Good ($>$80\%) & Excellent ($>$99\%)  \\
\hline
Angular Resolution  &  Good ($\approx$1-0.2 deg) & Excellent ($\approx$0.05 deg)  \\
\hline
Zenith Angle dependence  &  Very Strong ($\approx$cos$\theta^{-(6-7)}$) & Small ($\approx$cos$\theta^{-2.7}$)  \\
\hline
Effective Area  &  Shrinks with zenith angle & Increase with zenith angle \\
\hline
\hline
\end{tabular} 
\label{tab:one}
\end{table}

In shower arrays large areas (typically 10$^4$ - 10$^5$ m$^2$) are instrumented with charged particle detectors, usually scintillation counters, Resistive Plate Chambers (RPCs) or water Cherenkov tanks.
The tail particles of the shower are sampled at a single depth (the observational level) and at fixed distances from the shower core position. 
The key observables in all air shower arrays are the local shower particle densities and the secondary particle arrival times 
with which to reconstruct the shower arrival direction, the energy and the kind of the primary particle.
The resolutions of these measurements are limited by the large shower-to-shower fluctuations mainly due to the depth of the first interaction, which fluctuates by 1 radiation length ($\sim$37 g/cm$^2$) for electromagnetic particles and by 1 interaction length ($\sim$90 g/cm$^2$) for protons.

From an experimental point of view, the sampling of secondary particles at ground can be realized  with two different approaches
\begin{enumerate}
\item Particle Counting. A measurement is carried out with thin ($\ll$ 1 radiation length) counters providing a signal proportional to the number of charged particles (as an example, plastic scintillators or RPCs). The typical detection threshold is in the keV range.
\item Calorimetry. A signal proportional to the total incident energy of electromagnetic particles is collected by a thick (many radiation lengths) detector. An example is a detector constituted by many radiation lengths of water to exploit the Cherenkov emission of secondary shower particles. The Cherenkov threshold for electrons in water is 0.8 MeV and the light yield is $\approx$320 photons/cm or $\approx$160 photons/MeV emitted at 41$^\circ$.
\end{enumerate}

The mentioned experimental approaches have been applied in the last two decades in two different ground-based TeV survey instruments: 
\begin{enumerate}
\item[(1)] Water Cherenkov (Milagro) \cite{milagro}.
\item[(2)] Resistive Plate Chambers (ARGO-YBJ) \cite{disciascio-rev}. 
\end{enumerate}

In both experiments the instrumented areas of the core detectors were similar, a large central water reservoir 60$\times$80 m$^2$ for Milagro and a RPC carpet  74$\times$78 m$^2$ for ARGO-YBJ.
ARGO-YBJ was the first experiment to exploit the full coverage approach at extreme altitude.
The Water Cherenkov approach was later used in the HAWC and LHAASO experiments.
The RPC technology is widely used in particle physics experiments.

\section{The ARGO-YBJ experiment}
\begin{figure}[!t]
\centerline{\includegraphics[width=0.8\textwidth]{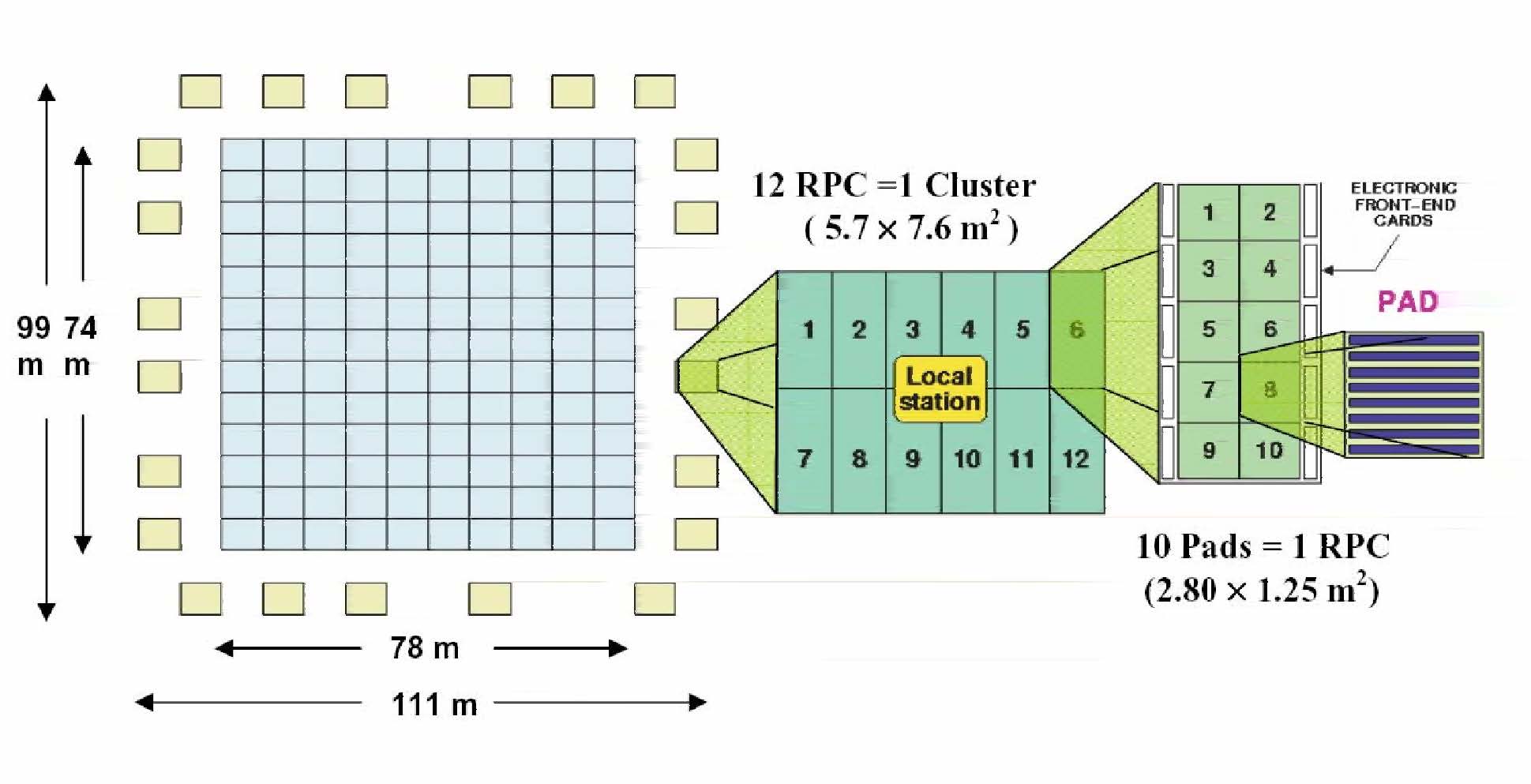}}
  \caption{Layout of the ARGO-YBJ experiment \cite{disciascio2017} (see text for a description of the detector).}
  \label{fig:fig-01}
\end{figure}

The ARGO-YBJ experiment, located at the Yangbajing Cosmic Ray Observatory (Tibet, PR China, 4300 m a.s.l., 606 g/cm$^2$), is constituted by a central carpet $\sim$74$\times$78 m$^2$, made of a single layer of Resistive Plate Chambers (RPCs) with $\sim$93$\%$ of active area, enclosed by a guard ring partially instrumented ($\sim$20$\%$) up to $\sim$100$\times$110 m$^2$. The apparatus has a modular structure, the basic data acquisition element being a cluster (5.7$\times$7.6 m$^2$), made of 12 RPCs (2.85$\times$1.23 m$^2$ each). Each chamber is read by 80 external strips of 6.75$\times$61.80 cm$^2$ (the spatial pixels), logically organized in 10 independent pads of 55.6$\times$61.8 cm$^2$ which represent the time pixels of the detector \cite{aielli06}. 
The readout of 18,360 pads and 146,880 strips is the experimental output of the detector. 
In addition, in order to extend the dynamical range up to PeV energies, each chamber is equipped with two large size pads (139$\times$123 cm$^2$) to collect the total charge developed by the particles hitting the detector \cite{argo-bigpad}.
The RPCs are operated in streamer mode by using a gas mixture (Ar 15\%, Isobutane 10\%, TetraFluoroEthane 75\%) for high altitude operation \cite{bacci00}. The high voltage settled at 7.2 kV ensures an overall efficiency of about 96\% \cite{aielli09a}.
The central carpet contains 130 clusters and the full detector is composed of 153 clusters for a total active surface of $\sim$6700 m$^2$ (Fig. \ref{fig:fig-01}). The total instrumented area is $\sim$11,000 m$^2$.
%
%

\begin{figure}[t!]
\begin{minipage}[t]{.47\linewidth}
  \centerline{\includegraphics[width=\textwidth]{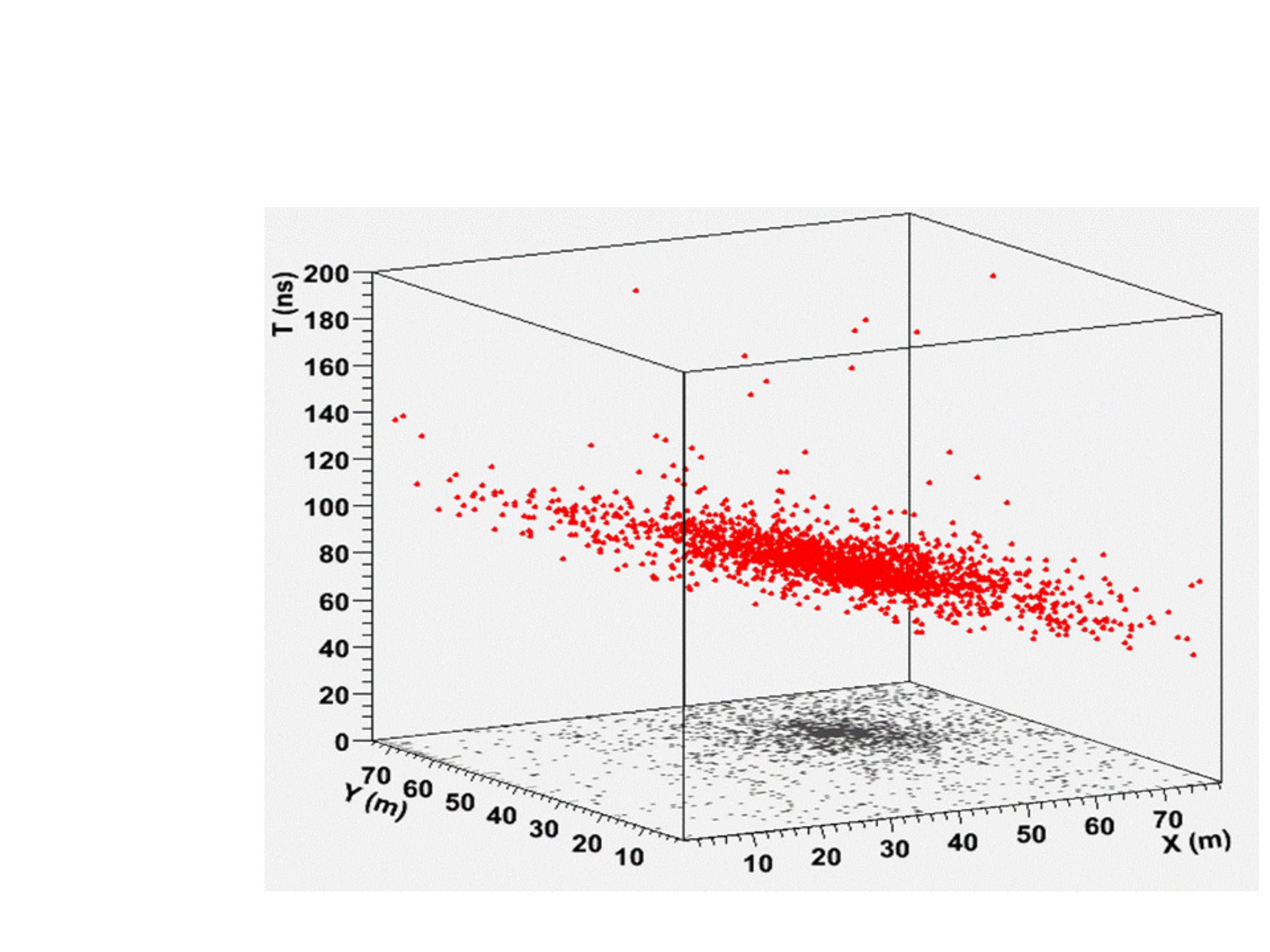} }
\end{minipage}\hfill
\begin{minipage}[t]{.47\linewidth}
  \centerline{\includegraphics[width=0.8\textwidth,angle=90]{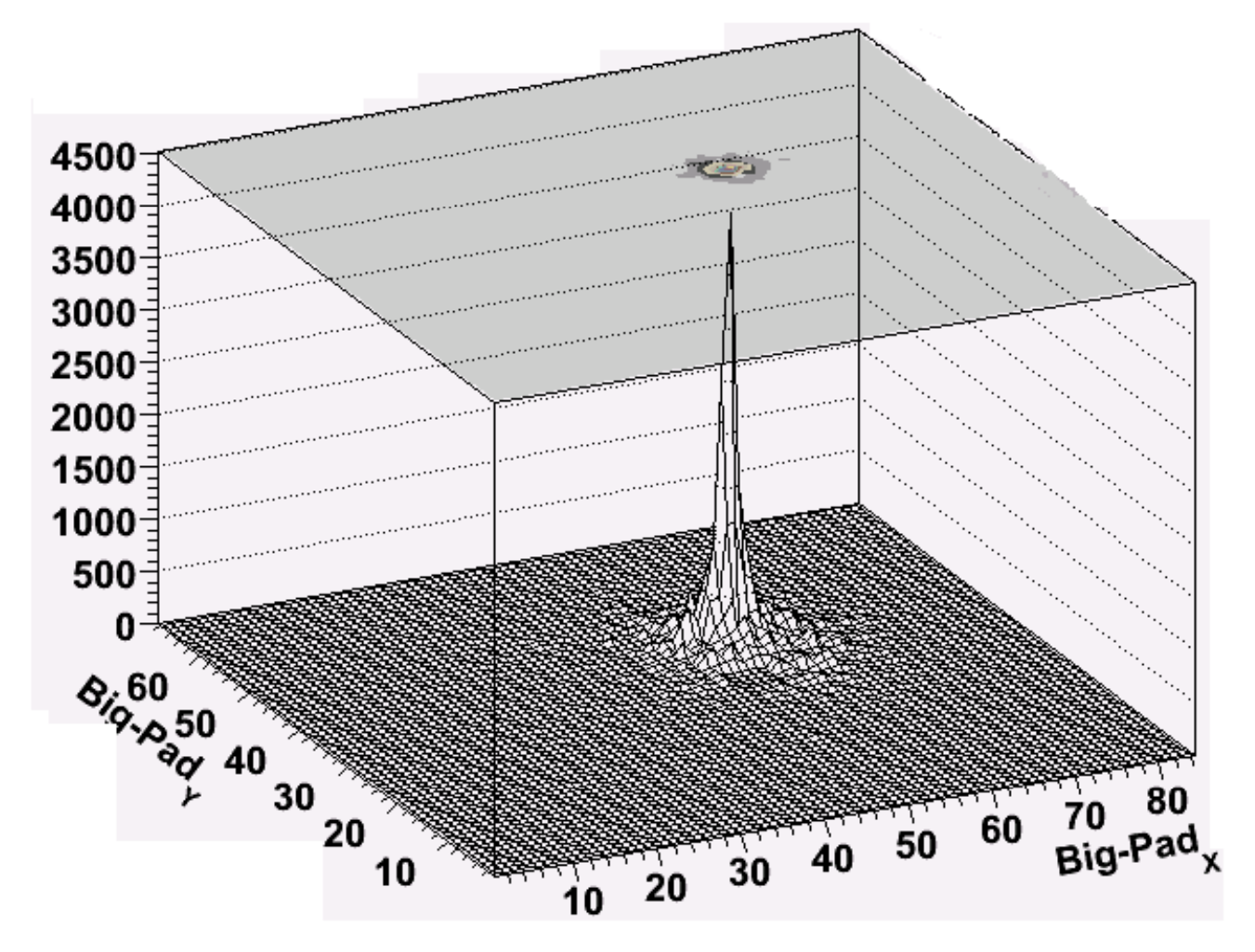} }
\end{minipage}\hfill
\caption[h]{Showers imaged by the ARGO-YBJ central carpet with the digital (left plot: the space-time structure of a low energy shower provided by the strip/pad system) and the charge (right plot: the core of a high energy shower imaged by the charge read-out) read-outs \cite{disciascio2017}.} 
\label{fig:argo-shower}
\end{figure}

Because of the small pixel size, the detector is able to record events with a particle density exceeding 0.003 particles m$^{-2}$, keeping good linearity up to a core density of about 15 particles m$^{-2}$.
This high granularity allows a complete and detailed three-dimensional reconstruction of the front of air showers at an energy threshold of a few hundred GeV, as can be appreciated in Fig. \ref{fig:argo-shower} (Left Plot) where a typical shower detected by ARGO-YBJ is shown. Showers induced by high energy primaries ($>$100 TeV) are also imaged by the charge readout of the large size pads (see Fig. \ref{fig:argo-shower}, Right Plot), which allow the study of shower core region with unprecedented resolution \cite{argo-bigpad}.

The median energy of the first multiplicity bin (20-40 fired pads) for photons with a Crab-like energy spectrum is $\sim$340 GeV \cite{argo-crab}, with a $\approx$50\% efficiency in the detection of 100 GeV gamma-induced showers.
The granularity of the read-out at centimeter level and a noise of accidental coincidences of 380 Hz/pad allowed to sample events with only 20 fired pads, out of 15,000, with a noise-free topological-based trigger logic.

The benefits in the use of RPCs in ARGO-YBJ are \cite{disciascio-rev,bartoli2011}: (1) high efficiency detection of low energy showers by means of the high density sampling of the central carpet (the detection efficiency of 100 GeV photon-induced events is $\approx$50\% in the first multiplicity bin); (2) unprecedented wide energy range investigated by means of the digital/charge read-outs ($\sim$300 GeV $\to$ 10 PeV); (3) good angular resolution ($\sigma_{\theta}\approx 1.66^{\circ}$ at the threshold, without any lead layer on top of the RPCs) and unprecedented details in the core region by means of the high granularity of the different read-outs.
RPCs allowed to study also charged CR physics (energy spectrum, elemental composition and anisotropy) up to about 10 PeV.

%
%

\subsection{The angular resolution}
 
The direction of the primary particle is obtained after reconstructing the time profile of the shower front by using the information from each timing pixel of the experiment. 
Shower particles are concentrated in a front of a nearly spherical shape. A good approximation for particles not far from the shower core is represented by a cone-like shape with an average cone slope of about 0.10 ns/m.
The accuracy in the reconstruction of the shower arrival direction mainly depends on the capability of measuring the relative arrival times of the shower particles. 

The angular resolution of a shower array is a combination of the temporal resolution of the detector unit, the dimension of apparatus, i.e. the dimension of the lever arm in the fitting procedure of the shower front, and the number of temporal hits, i.e. the granularity of the sampling.

%
\begin{figure}[h!]
\begin{minipage}[t]{.47\linewidth}
  \centering
  \includegraphics[width=\textwidth]{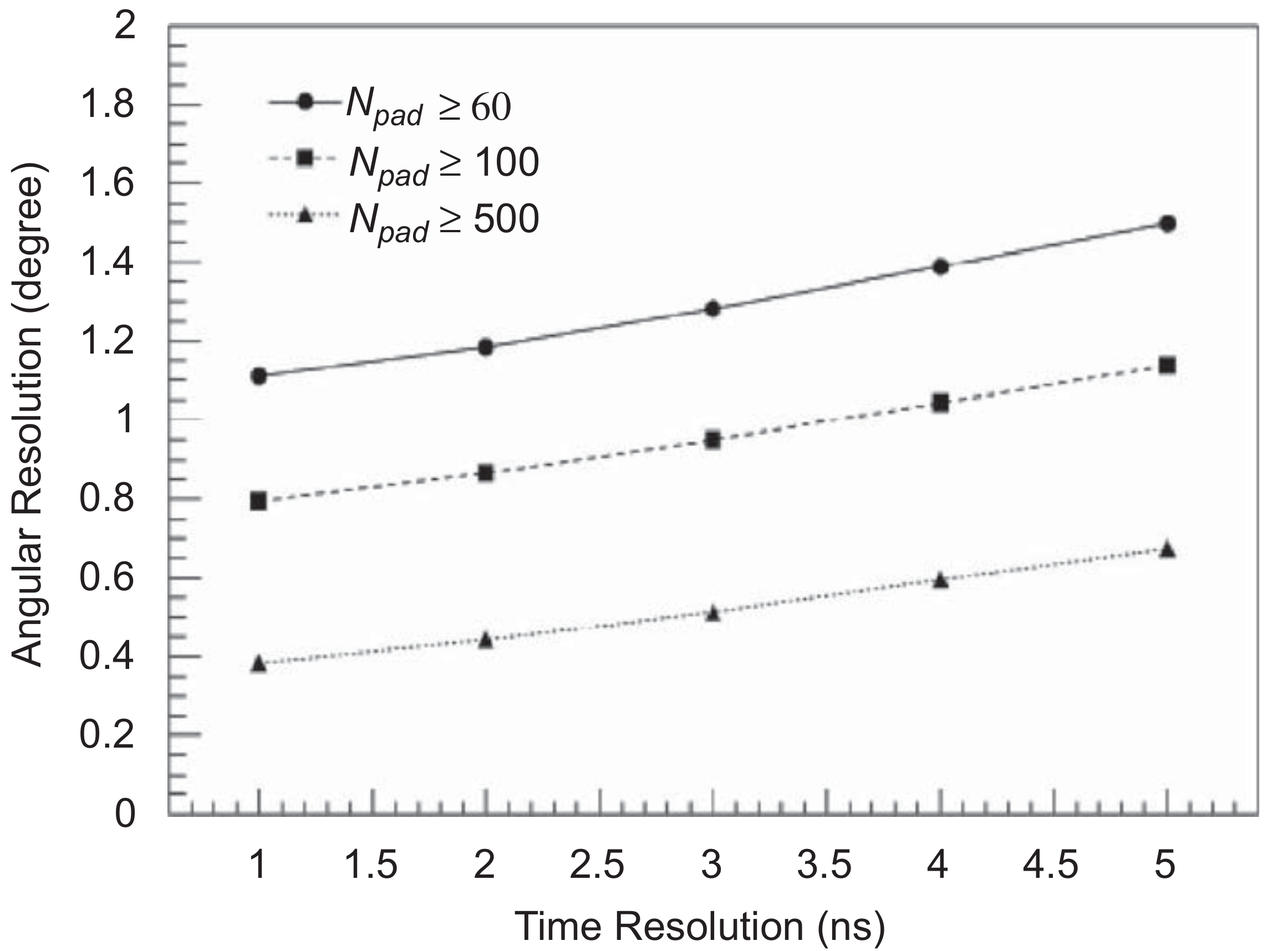} 
\end{minipage}\hfill
\begin{minipage}[t]{.48\linewidth}
  \centering
  \includegraphics[width=\textwidth]{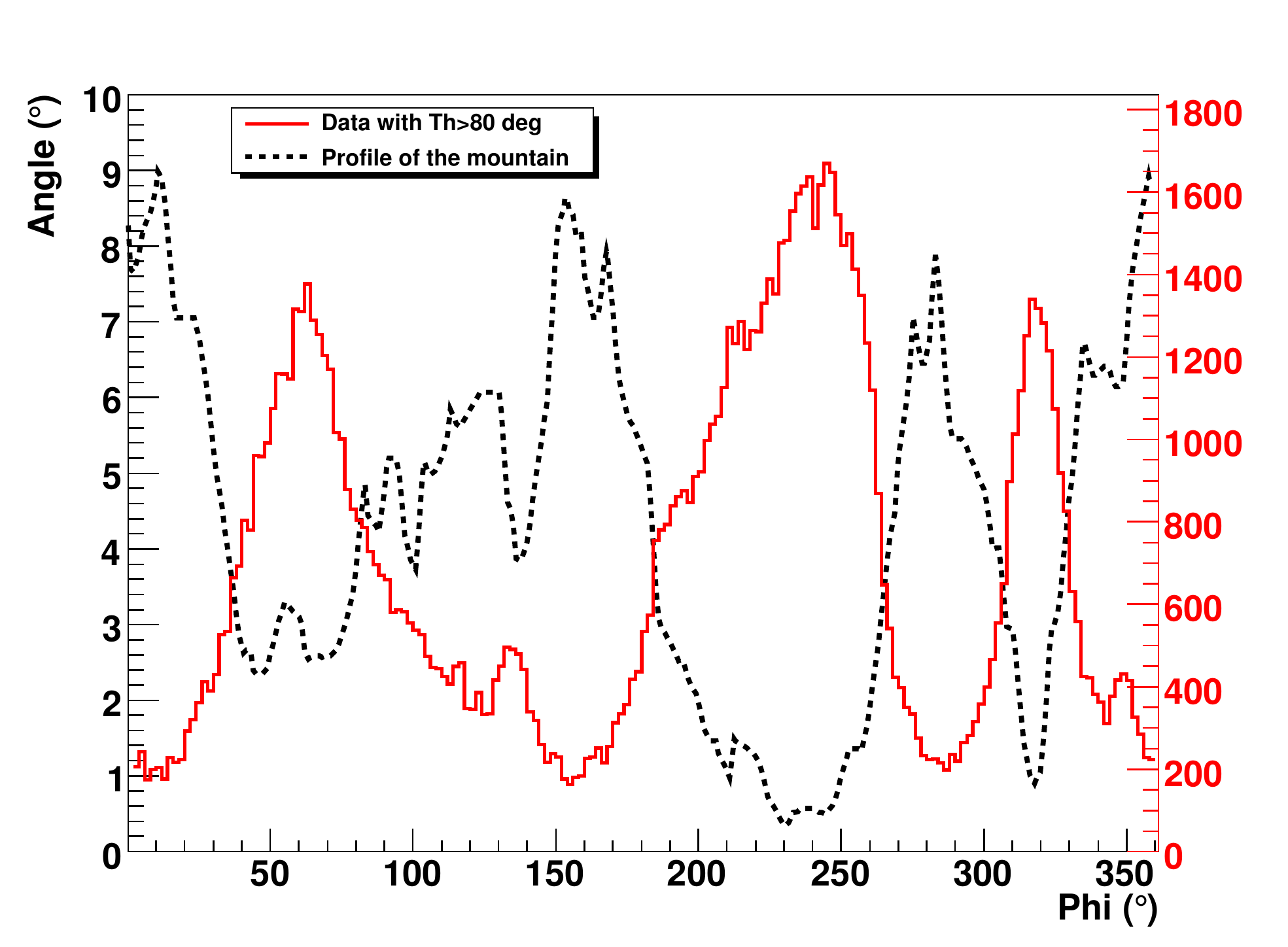} 
\end{minipage}\hfill
\caption{Left Panel: Angular resolution vs time resolutions for RPCs in the ARGO-YBJ experiment. Events with N$_{pad}\geq$60, 100 and 500 fired pads have been selected \cite{aielli2009}. Right Panel: Azimuthal distribution of showers arriving with a zenith angle $>80$ deg compared with the mountain landscape around the detector.}
\label{fig:angres}
\end{figure}
%

The time resolution of each detector is determined by the intrinsic time resolution, the propagation time of the signal and the electronic time resolution. As an example, for the ARGO-YBJ experiment the total detector resolution is $\approx$1.3 ns (including RPC intrinsic jitter, strip length, electronics time resolution) \cite{bartoli2011}.
The dependence of the angular resolution on the time resolution of RPCs in the ARGO-YBJ experiment is shown in the Left Panel of Fig. \ref{fig:angres} \cite{aielli2009}. Events with N$_{pad}\geq$60, 100 and 500 fired pads on the central carpet have been selected.
As it can be seen from the figure, a time resolution in the range between 1 and 2 ns corresponds to a very small change in the angular resolution because the time jitter of the earliest particles in high multiplicity events (>100 hits) is estimated $\approx$1 ns \cite{epas2,argo-test}.

The ARGO-YBJ capability in reconstructing the shower direction is showed in Right Panel of Fig. \ref{fig:angres} where the azimuthal distribution of showers arriving with a zenith angle $>80$ deg is compared with the mountain landscape around the detector. As it can be seen, an impressive strong anti-correlation between shower flux and mountains is clearly evident, even for narrow valleys between the mountains. We emphasize that no lead layer to improve the angular resolution by converting secondary photons and absorbing delayed electrons is added on top of RPCs. Therefore, this plot shows the shower directtion reconstruction capability of RPCs alone.

\section{Cosmic Ray background rejection from ground}

In 1960 Maze and Zawadzki \cite{maze1960} suggested that in gamma-ray astronomy with shower arrays the background of CRs can be identified and rejected by identifying EAS with an abnormally small number of muons N$_{\mu}$, the so-called \emph{'muon poor'} technique. The existence of such 'unusual' showers is due to the relatively small photo-nuclear cross section compared with the corresponding value for the proton-nucleus and nucleus-nucleus cross-sections. In gamma showers muons are produced mainly by the photo-production of hadrons $\gamma$ + air $\to$ n$\pi^{\pm}$ + m$\pi^0$ + X ($\sigma_{\gamma-air}\sim$ 1-2 mb), followed by the pion decays in muons and photons, and by muon pair production ($\sigma_{\gamma-air}\sim$12 $\mu$b).

The efficacy of background rejection exploiting the muon content is limited by the number of muons that can be detected. 
According to MonteCarlo simulations, in a proton-induced shower the number of muons is approximately proportional to the
energy of the primary, with about 20 muons above 1 GeV for a 1 TeV proton (200 muons for a 10 TeV muon), but only 4 muons within 150 m of the shower core. As a consequence, the muon poor technique is effective above a few TeV. 

In addition, the fluctuations in the muon number (for a fixed proton energy) are larger than Poisson, with a Gaussian width of $\approx 2.5\sqrt{N_\mu}$, thus there are more events with zero muons than a Poisson calculation. This is an important limiting factor for background discrimination at low energy. Another limiting factor is the high rate of single muons unassociated to showers at ground. 
The need for \emph{large full coverage muon detector} is evident to exploit the muon poor technique starting from the TeV energy range.

The background-free regime is very important because in this case the sensitivity is the inverse of the effective area of the array multiplied by the time spent observing a source. Thus, an EAS array with a comparable effective area to a IACT array, with more than one order of magnitude larger time on source, will have a much better sensitivity to highest energy sources.

ARGO-YBJ was not able to discriminate muons, anyway, we note that in the Milagro and ARGO-YBJ experiments the limited capability to discriminate the background was mainly due to the small dimensions of the central detectors (pond and carpet). In fact, in the new experiments HAWC \cite{hawc1} and LHAASO \cite{lhaaso1,cao2021nat}, the discrimination of the CR background is made studying shower characteristics far from the shower core (at distances R$>$ 40 m from the core position, the dimension of the Milagro pond and ARGO-YBJ carpet). 

\section{What's next? STACEX in the Southern Hemisphere}
\label{sec:stacex}

The recent results obtained by the LHAASO Collaboration with half detector in data taking revealed the existence of a large number of gamma-ray sources emitting photons with energies well beyond 500 TeV \cite{cao2021nat}.
The unexpected observation of these sources in the Northern hemisphere suggests the opportunity to discover tens of similar Ultra High Energy emissions in the Inner Galaxy by a detector able to detect PeV photons located in the Southern hemisphere.

An ideal observatory for PeVatrons
\begin{itemize}
\item should be able to perform an unbiased survey to search for different and possibly unexpected classes of sources;
\item should have a dynamical range from 100 GeV to 10 PeV to measure the energy spectra with the same detector;
\item should have an effective area bigger than 0.5 km$^2$ to collect adequate photon statistics;
\item should have a very good energy resolution (20\% or better) above a few tens of TeV to test spectral break and cutoffs;
\item should have a good angular resolution ($\sim 0.2^{\circ}$) to resolve sources which might be hidden in the tails of bright sources and compare and correlate with gas surveys;
\item should have a background discrimination capability at a level of 10$^{-4}$ - 10$^{-5}$ starting from a few tens of TeV.
\end{itemize}

\begin{figure}[t!]
\begin{minipage}[t]{.47\linewidth}
  \centerline{\includegraphics[width=\textwidth]{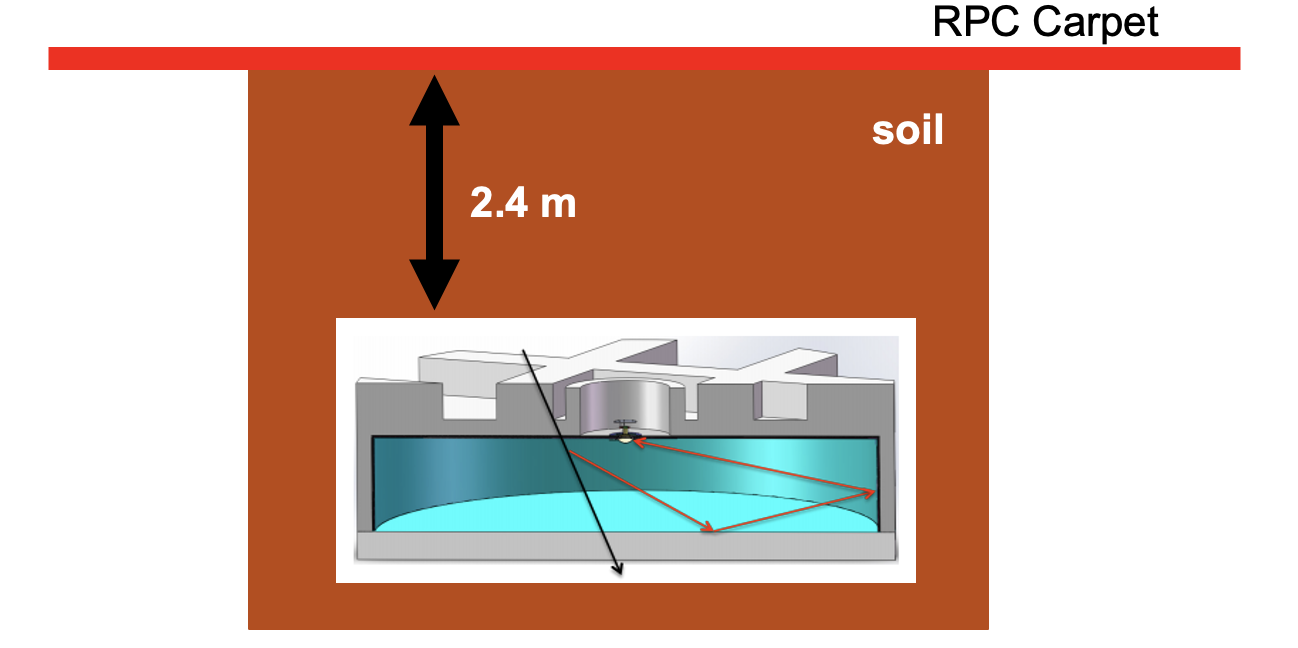} }
\end{minipage}\hfill
\begin{minipage}[t]{.47\linewidth}
  \centerline{\includegraphics[width=\textwidth]{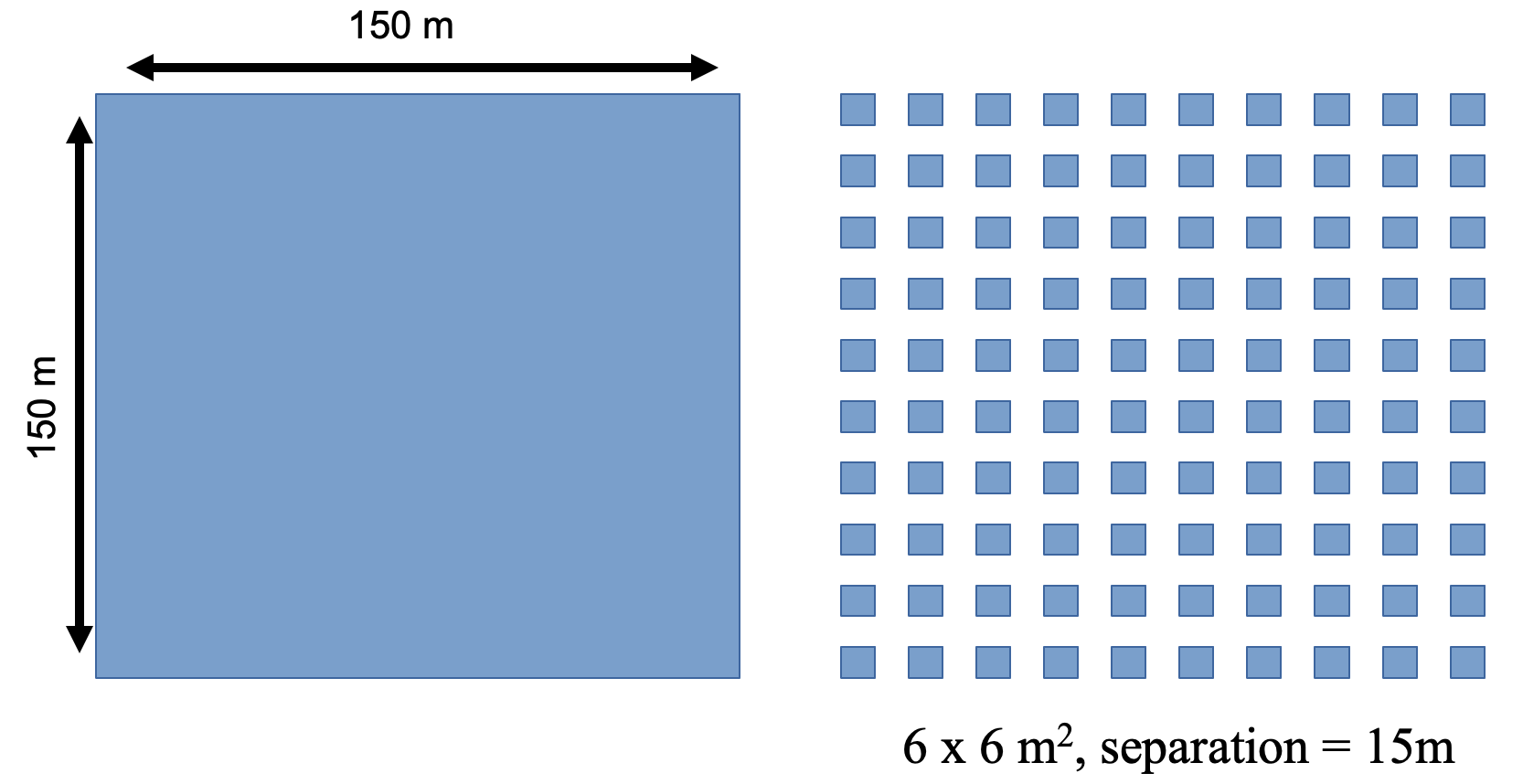} }
\end{minipage}\hfill
\caption[h]{Left Panel: STACEX layout. Right Panel:  Layouts of 2 different muon detectors under the RPC carpet.} 
\label{fig:stacex}
\end{figure}

The proposed STACEX (Southern TeV Astrophysics and Cosmic rays EXperiment) detector (see Fig. \ref{fig:stacex}) \cite{stacex-icrc2021} consists of a full coverage core detector, located at 5000 m asl, constituted by
\begin{itemize}
\item[(a)] a 150$\times$150 m$^2$ RPC full coverage carpet, with a 0.5 mm lead layer on top;
\item[(b)] a dense muon detector array below the carpet constituted by water Cherenkov tanks buried under 2.5 m of soil;
\end{itemize}

Adequate photon statistics above 50 TeV is provided by an array of 1 m$^2$ plastic scintillators and muon detectors in a 30 m grid  around the core detector covering a total area of 0.5 km$^2$ at least. 

This layout is motivated by following reasons
\begin{itemize}
\item[(1)] dense sampling by the RPC carpet for a very low energy threshold ($\sim$100 GeV);
\item[(2)] wide energy range, 100 GeV $\to$ 10 PeV;
\item[(3)] high granularity of the carpet read-out to have an energy resolution $<$20\% above 10 TeV and to have a very good angular resolution ($\sim 0.2^{\circ}$ above 10 TeV);
\item[(4)] high efficiency rejection of the CR background by the muon-poor technique. With a continuous 22,500 m$^2$ muon detector below the RPC carpet the background-free detection of gamma-rays is expected to start from a few tens TeV;
\item[(5)]  measurement of the elemental composition with two different independent observables, shower core characteristics (in a ARGO-like measurement \cite{disciascio-rev}) and muon component.
\end{itemize}

%
%

\begin{figure}[t!]
\begin{minipage}[t]{.47\linewidth}
  \centerline{\includegraphics[width=\textwidth]{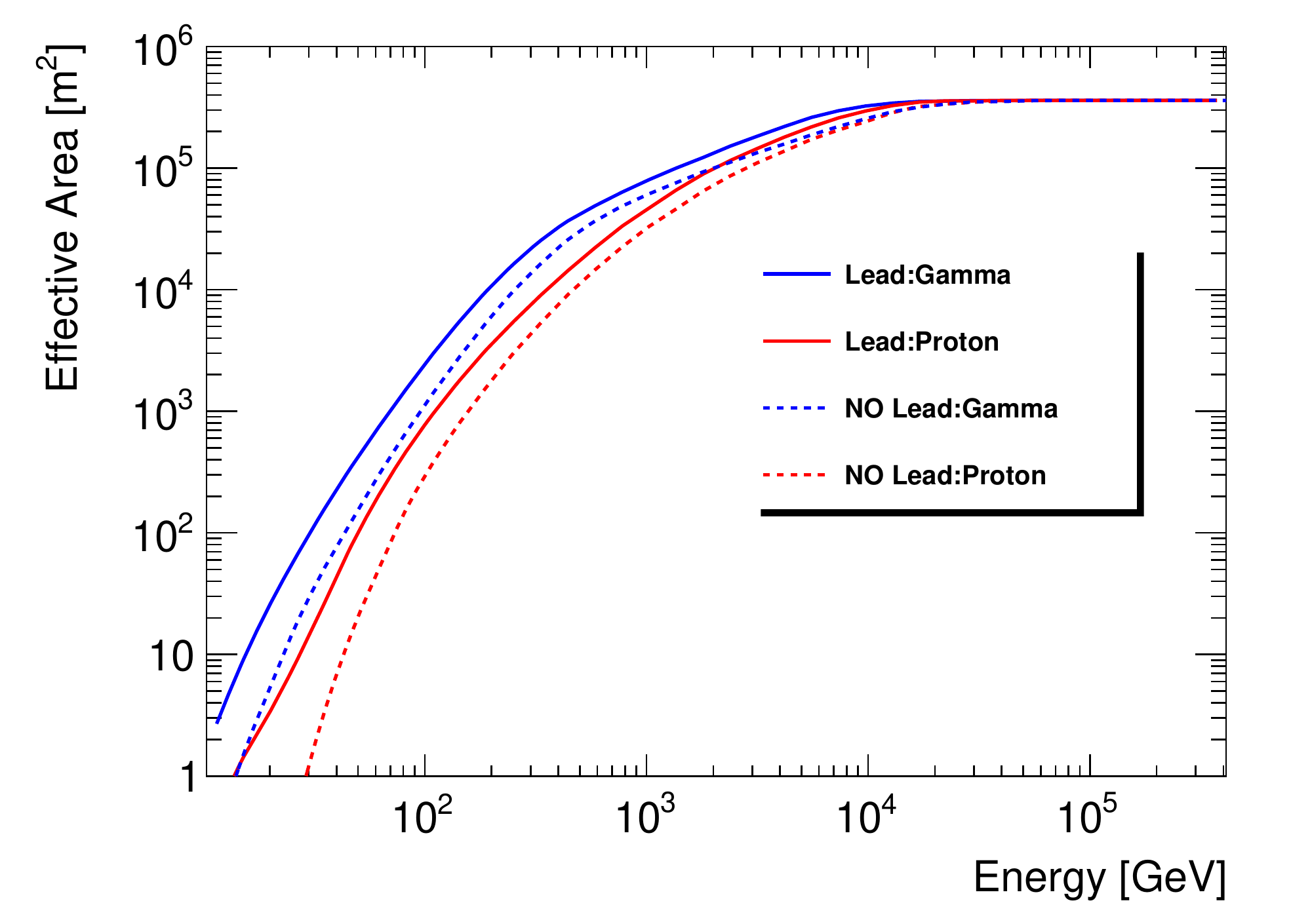} }
\end{minipage}\hfill
\begin{minipage}[t]{.47\linewidth}
  \centerline{\includegraphics[width=\textwidth]{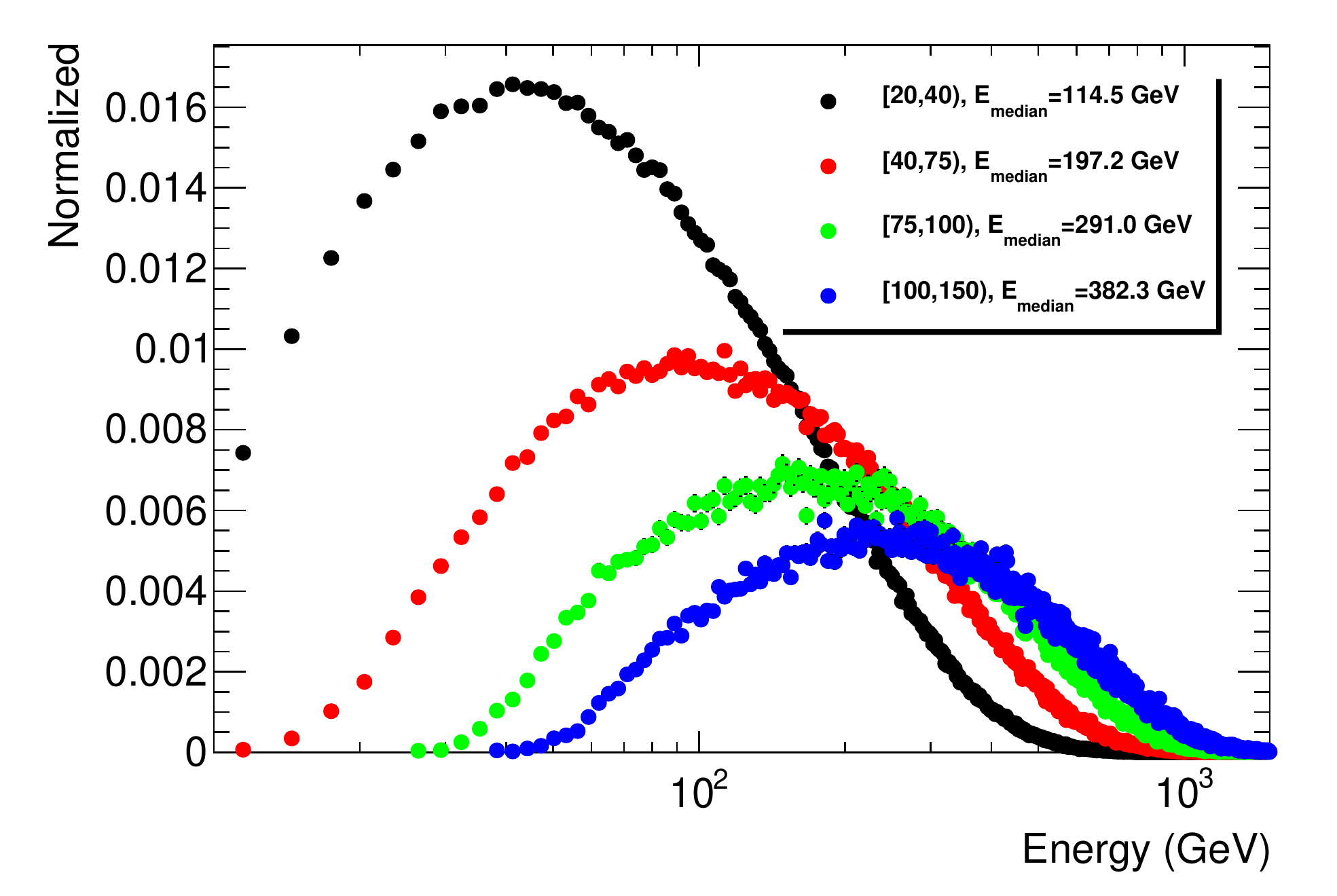} }
\end{minipage}\hfill
\caption[h]{Left Panel: STACEX effective areas for photon- and proton-induced showers as a function of the primary energy. Right Panel:  Energy distributions of photon-induced showers for 4 different strip multiplicities measured by the RPC carpet \cite{stacex-icrc2021}.} 
\label{fig:areaef_medianen}
\end{figure}

Results of preliminary calculations are shown in the following \cite{stacex-icrc2021}.
The effective area for showers produced by primary photons and protons are shown in the Left Panel of Fig. \ref{fig:areaef_medianen} as a function of the median energy for different bins of strips multiplicity. As you can see from the figure, we have A$_\text{eff}\sim$ 3$\times$10$^3$ m$^2$ at 100 GeV and A$_\text{eff}\sim 10^6$ m$^2$ at 100 TeV. 
The energy distributions are shown in the Right Panel of Fig. \ref{fig:areaef_medianen} for 4 different strip multiplicities. The peak energy of the first bin is about 100 GeV. The energy resolution is about 50$\%$ at the threshold.

\begin{figure}[t!]
\begin{minipage}[t]{.47\linewidth}
  \centerline{\includegraphics[width=\textwidth]{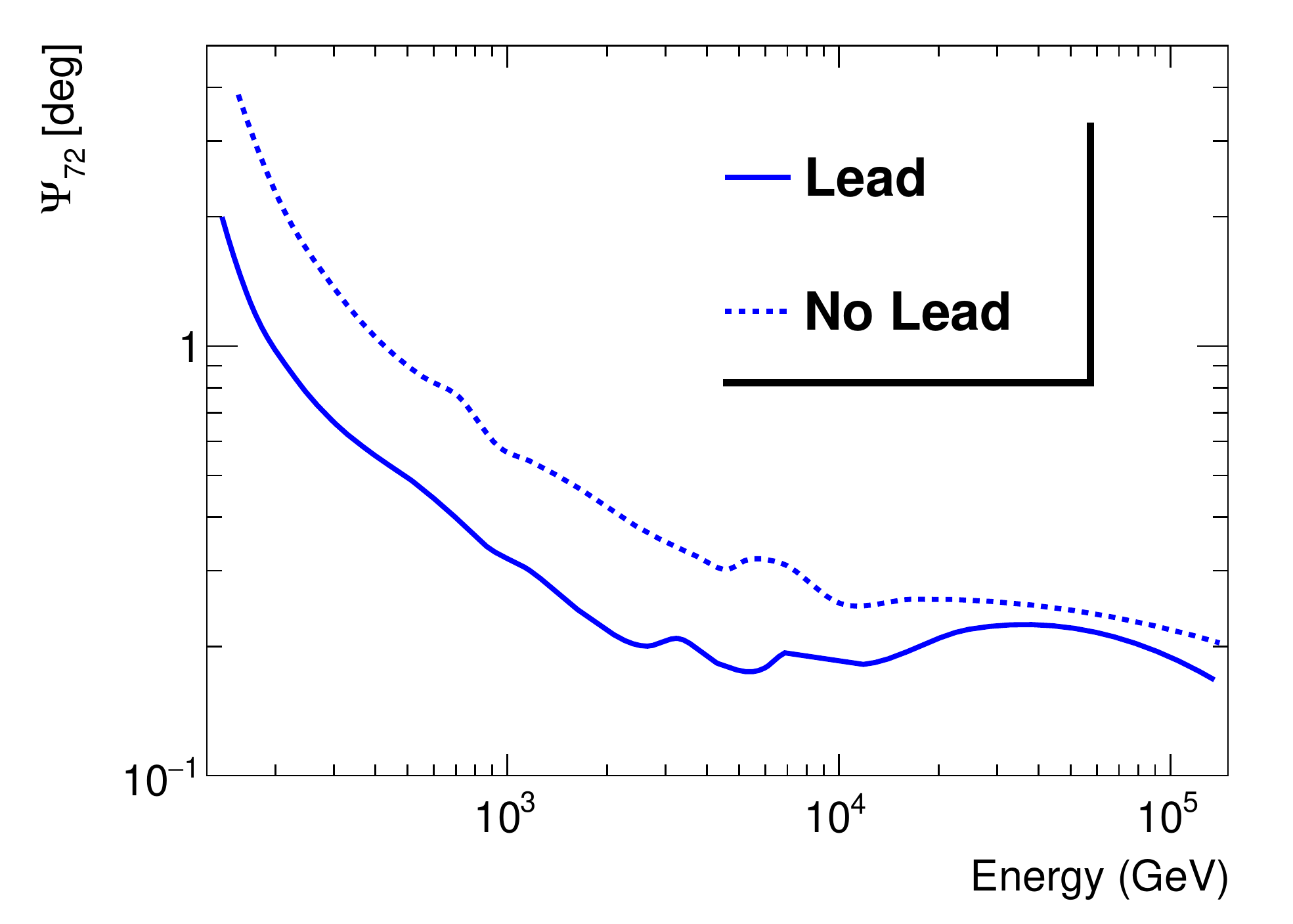} }
\end{minipage}\hfill
\begin{minipage}[t]{.47\linewidth}
  \centerline{\includegraphics[width=\textwidth]{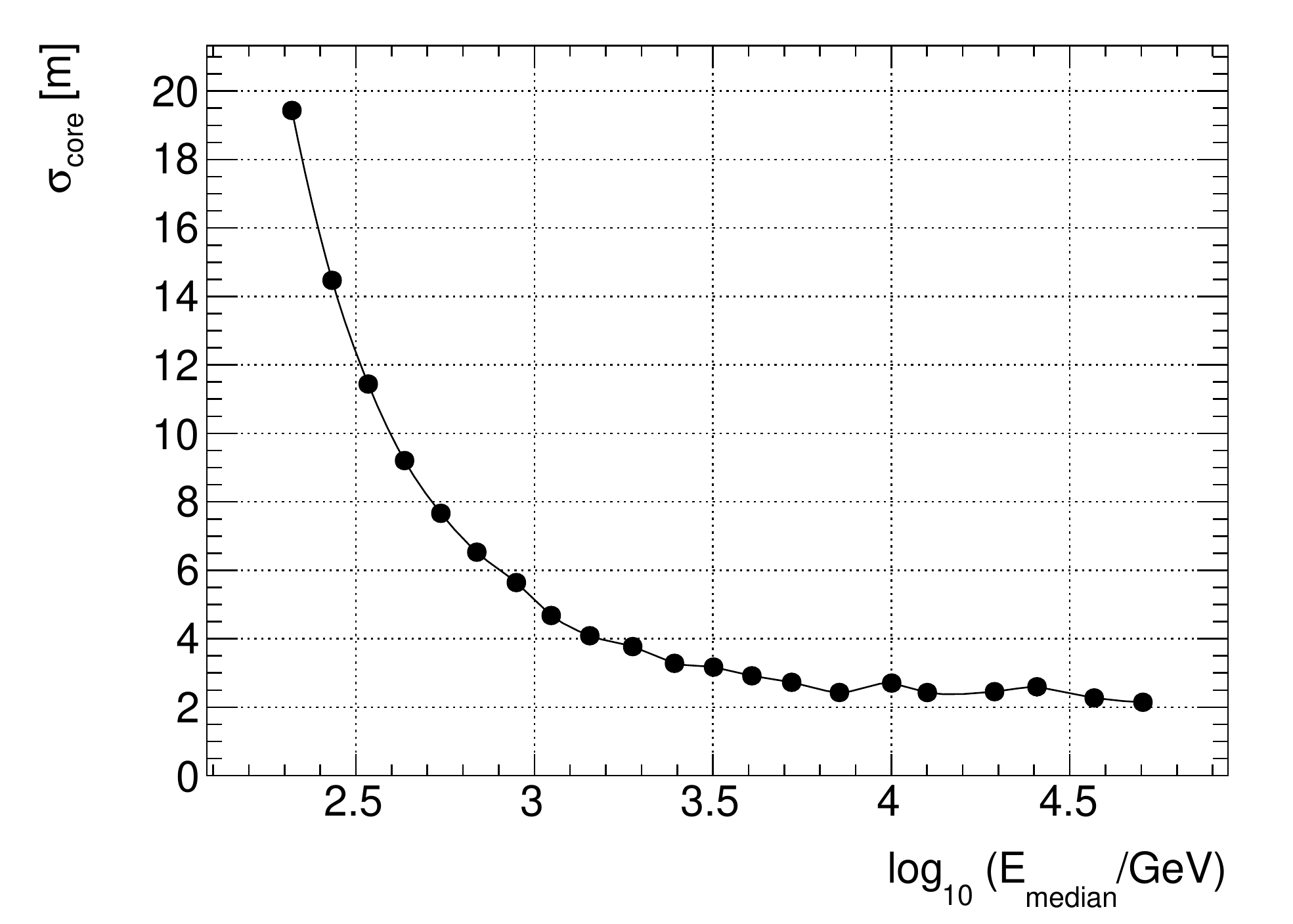} }
\end{minipage}\hfill
\caption[h]{Left Panel: Angular resolution for photon-induced showers as a function of the primary energy. Right Panel: Shower core resolution for photon-induced showers as a function of the primary energy \cite{stacex-icrc2021}.} 
\label{fig:angres_coreres} 
\end{figure}

In the Left Panel of Fig. \ref{fig:angres_coreres} the angular resolution $\sigma_{\theta}$ (the angle containing the 72\% of the events) is shown. We have $\sigma_{\theta}\sim$0.5$^{\circ}$ at 1 TeV  and $\sigma_{\theta}\sim$0.25$^{\circ}$ at 10 TeV. The core resolution for gamma-ray events is shown in the Right Panel of Figure \ref{fig:angres_coreres} as a function of the reconstructed energy. The resolution is about 20 m at 100 GeV and $\sim$2 m at 100 TeV.

%
%

%
%

\subsection{Full coverage detection of muons}

The lateral distribution of muons is much wider than electromagnetic particles and their number much lower. Therefore, the detection of muons can be affected by large sampling fluctuations. Detection of muons with a full coverage approach can lower the fluctuations and allow a background-free measurement in UHE gamma-ray astronomy starting from a few tens of TeV.

To evaluate the background discrimination capability of the proposed array we simulated 2 different water Cherenkov muon detector layouts: (1) a continuous muon detector below the RPC carpet with a total area of 22,500 m$^2$; (2) a 10$\times$10 array of water tanks with a total area of 3,600 m$^2$.
In both cases the detectors, made by 1.5 m of water with downward PMTs, are buried under 2.5 m of soil to reduce the punch-through probability by high energy electromagnetic particles. As a consequence, the muon energy threshold is about 1 GeV.
To further reduce the contamination in the analysis we excluded the muon detectors inside a circular area with 20 m radius around the reconstructed shower core position. 

\begin{figure}[h]
\begin{subfigure}{0.33\textwidth}
\includegraphics[width=\textwidth]{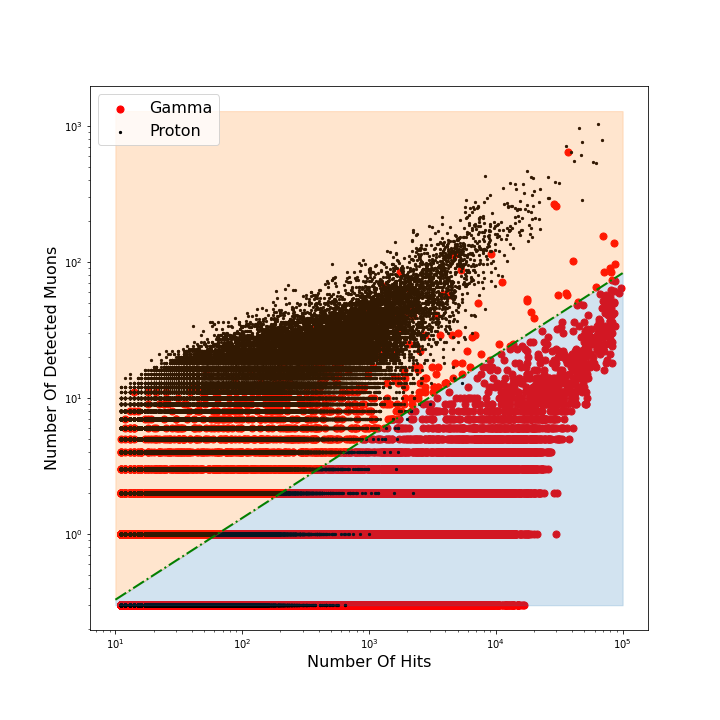} 
\end{subfigure}
\begin{subfigure}{0.33\textwidth}
\includegraphics[width=\textwidth]{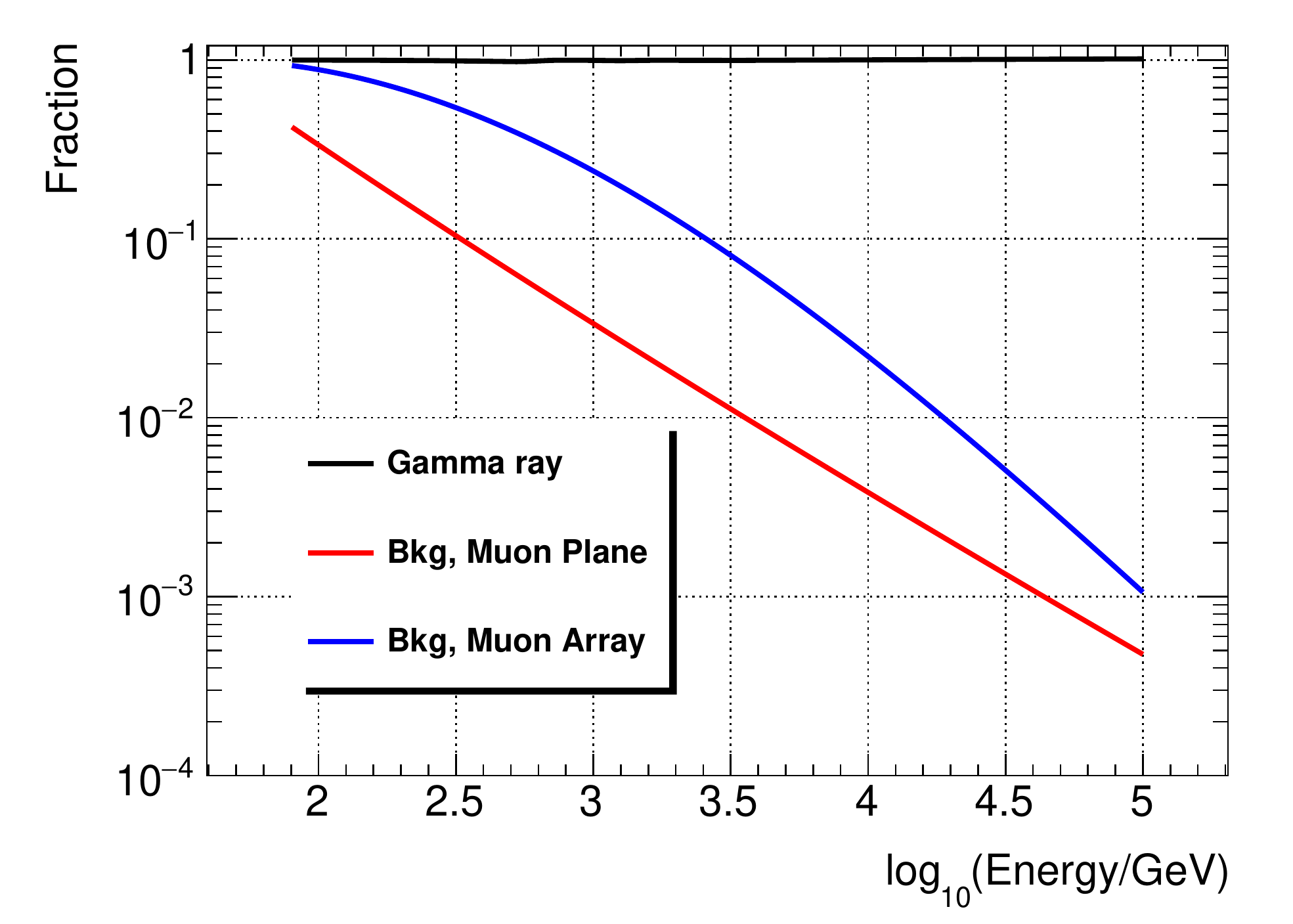} 
\end{subfigure}
\begin{subfigure}{0.33\textwidth}
\includegraphics[width=\textwidth]{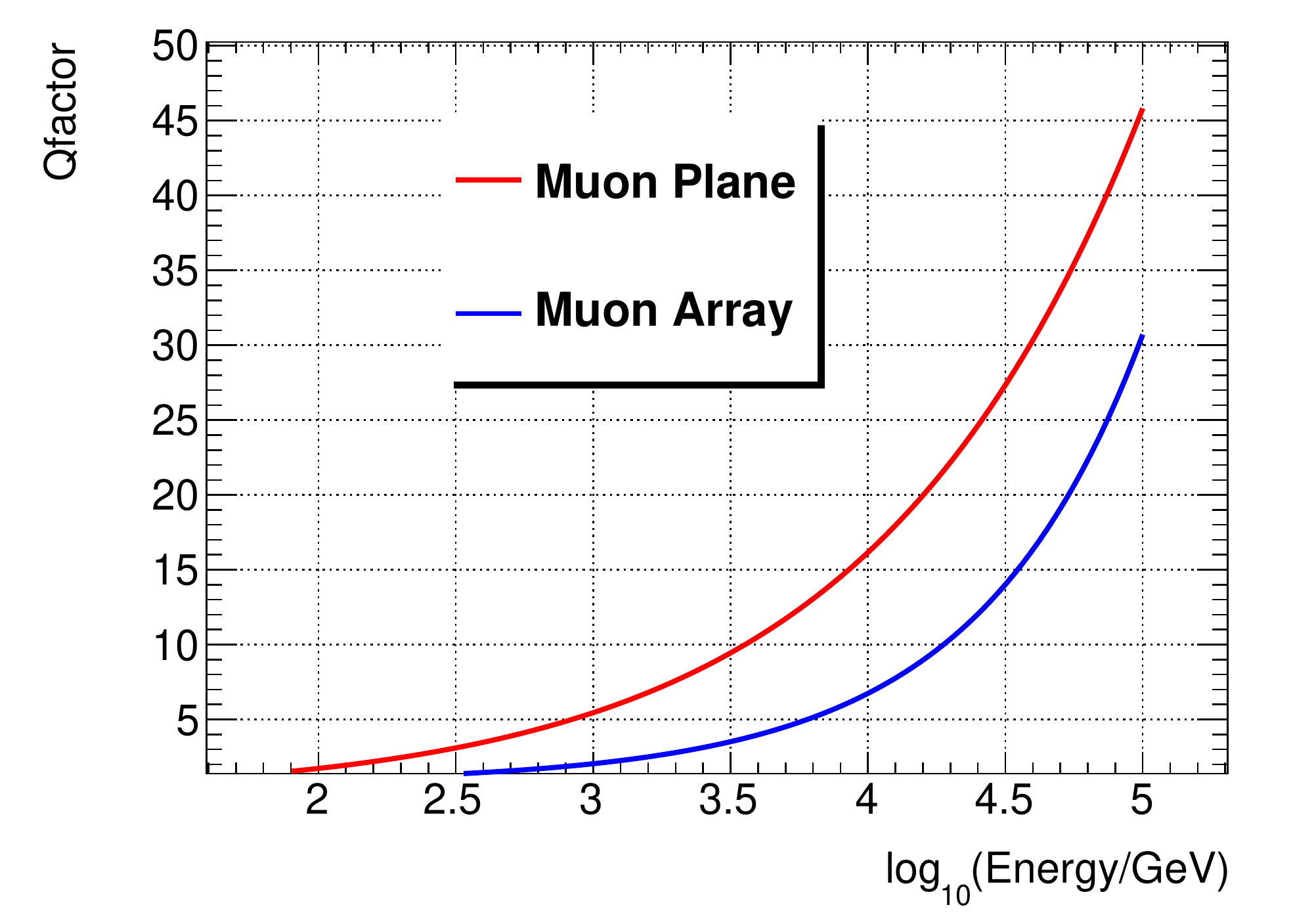} 
\end{subfigure}
\caption{Left panel: Number of muons detected by a 150$\times$150 m$^2$ continuous muon detector for proton and photon induced showers as a function of the strip multiplicity. Middle panel: The survival fraction of gamma-ray and cosmic ray background events as a function of the energy. Right panel: Q-factor parameter as a function of the photon primary energy for the 2 investigated muon detector layouts \cite{stacex-icrc2021}.}
\label{fig:discrimination}
\end{figure}

In the Left Panel of Figure  \ref{fig:discrimination} the number of muons detected by a 150$\times$150 m$^2$ continuous muon detector for proton- and photon-induced showers is shown as a function of the strip multiplicity. 
We reject the CR background according to a selection cut removing showers with a muon content bigger than a value determined to optimize the sensitivity as a function of the multiplicity.
The fraction of the photons and protons surviving a selection cut determined using a binary classification method with a logistic function as a function of the primary photon energy is shown in the Middle Panel of Figure \ref{fig:discrimination}. For energies above 100 TeV  we reject the proton background at a level of 3$\cdot$ 10$^{-4}$ with nearly 100$\%$ of photons surviving. The so-called \emph{'Q-factor'} parameter is shown in the Right Panel as a function of the primary energy for the 2 investigated muon detector layouts. 
Calculations are still preliminary but these results show that with a full coverage measurement of muons the background-free regime could start at a few tens of TeV.

%
%

\begin{figure}[t!]
  \centerline{\includegraphics[width=\textwidth]{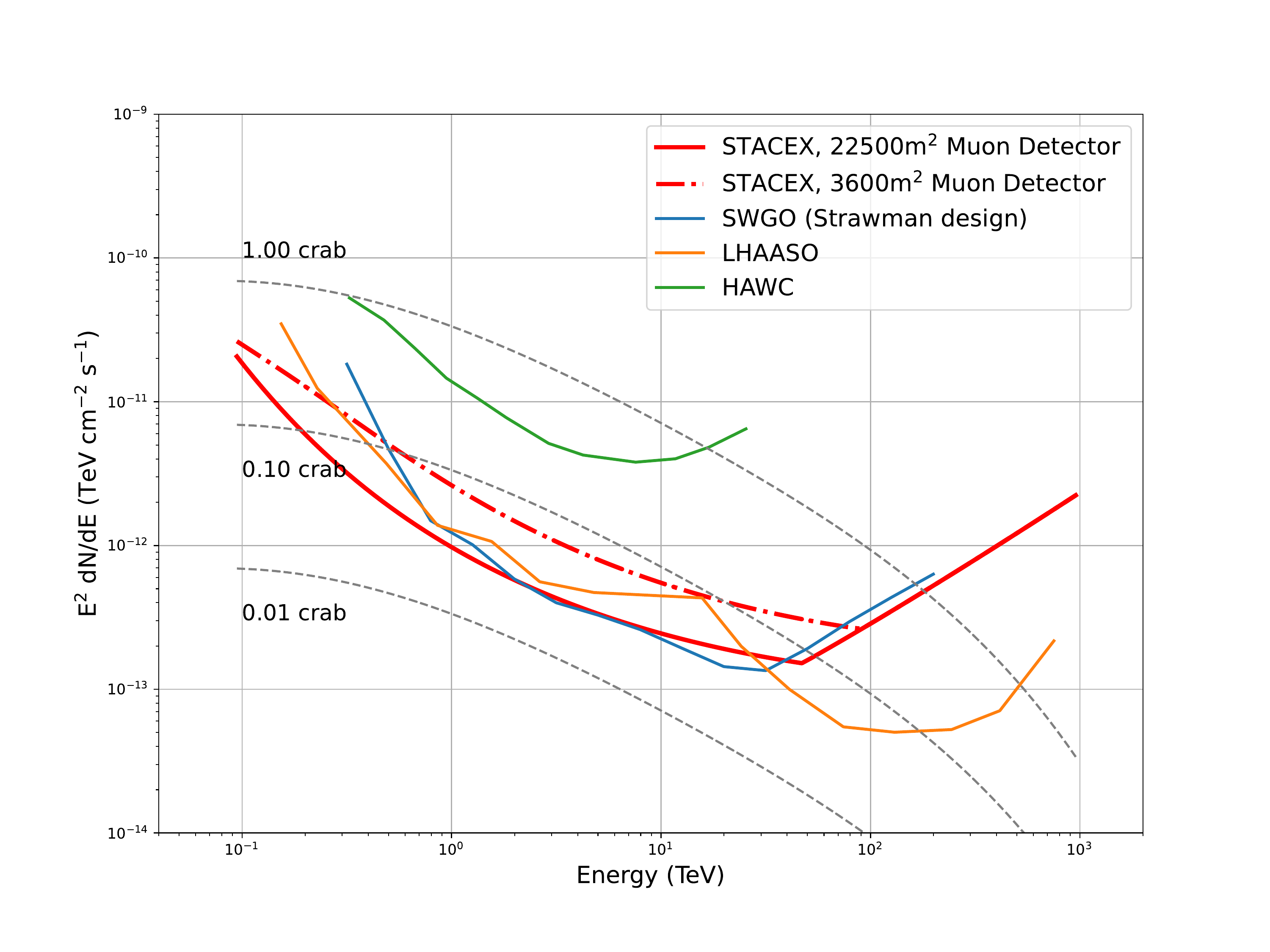} }
\caption[h]{Differential sensitivity of STACEX central detector in 1 year compared to HAWC (22,500 m$^2$ at 4100 m asl) \cite{hawc1}, LHAASO (1.3 km$^2$ at 4400 m asl) \cite{lhaaso1}  and SWGO (220,000 m$^2$ at 5000 m asl) \cite{swgo-2019}. The calculations refer to the 2 different muon detector layouts discussed in the text. The effect of a shower array located around the STACEX central carpet is not included yet \cite{stacex-icrc2021}.}
\label{fig:sensitivity} 
\end{figure}

In Fig. \ref{fig:sensitivity} the sensitivity of the STACEX central detector, located at 5000 m asl \cite{stacex-icrc2021}, is compared to that of LHAASO at 4400 m asl \cite{lhaaso1} and SWGO at 5000 m asl \cite{swgo-2019}.

As you can see, a smaller full coverage detector has a sensitivity comparable with much larger arrays like LHAASO or SWGO. The main reason is the full coverage layout of the muon detector allowing to efficiently reject charged cosmic ray background in the TeV range and to have a background-free regime starting from about 50 TeV.
We remind that in LHAASO the large area muon detector consists in a sparse array (coverage $\approx
 4.4\times10^{-2}$) with an energy threshold of about 10 TeV.
 In addition the full coverage of the RPC carpet allows a very low energy threshold.
The effect of a shower array located around the central carpet is not included yet.

%
%

\section{Conclusion}
Detection of EAS with a full coverage detector is the right solution to lower the energy threshold in the 100 GeV range, to study the characteristics of showers with unprecedented details and to discriminate the background of charged CRs working in a background-free regime starting from a few tens of TeV.

ARGO-YBJ exploited this approach operating for the first time a full coverage carpet at 4300 m asl showing that bakelite-based RPCs can be safely operated at extreme altitudes for many years providing: (1) high efficiency detection of low energy showers (energy threshold $\sim$100 GeV) by means of the dense sampling of the full coverage carpet; (2) unprecedented wide energy range investigated by means of the digital/charge read-outs ($\sim$100 GeV $\to$ 10 PeV); (3) good energy and angular resolutions with unprecedented details in the shower core region by means of the high granularity of the read-outs.

Coupled with a full coverage muon detector this apparatus should be able to detect photons in a background-free regime starting from a few tens of TeV.

\footnotesize{

}

\nolinenumbers

\begin{thebibliography}{99}

\providecommand{\eprint}[2][]{\url{#2}}


\bibitem{milagro}
 R. Atkins et al., \nima \textbf{449}, 478 (2000).
\bibitem{disciascio-rev}
G. Di~Sciascio {\em Int. J. Mod. Phys.\/} {\bf D23} 1430019 (2014) [Erratum: Int.
  J. Mod. Phys. D24, no.02, 1592001 (2014)]
  \bibitem{disciascio2017}
 G. Di Sciascio, {\em J. of Phys.: Conf. Series} {\bf 866}, 012017 (2017).
\bibitem{aielli06} G. Aielli \etal (ARGO-YBJ Coll.), \nima \textbf{562}, 92 (2006).
\bibitem{argo-bigpad} B. Bartoli \etal (ARGO-YBJ Coll.), \app \textbf{67}, 47 (2015).
\bibitem{bacci00} C. Bacci \etal (ARGO-YBJ Coll.), \nima \textbf{443}, 342 (2000).
\bibitem{aielli09a} G. Aielli \etal (ARGO-YBJ Coll.), \nima \textbf{608}, 246 (2009).
\bibitem{argo-crab}
B. Bartoli \etal (ARGO-YBJ Coll.) {\em Astrophys. J.\/} {\bf 798} 119 (2015).
  (\textit{Preprint} \eprint{1502.05665})
\bibitem{bartoli2011} 
B. Bartoli, B. \etal (ARGO-YBJ Coll.) {\em Phys. Rev. } {\bf D84}, 022003 (2011).
\bibitem{aielli2009}
G. Aielli \etal (ARGO-YBJ Coll.) NIM {\bf A608}, 246 (2009).
\bibitem{epas2}
G. Di Sciascio, B. D'Ettorre Piazzoli and M. Iacovacci, Astropart. Phys. {\bf 6}, 313 (1997).
\bibitem{argo-test}
C. Bacci \etal, Astroparticle Physics {\bf 17}, 151 (2002).
\bibitem{maze1960} 
R. Maze and A. Zawadzki, Nuovo Cimento {\bf 17}, 625 (1960).
\bibitem{hawc1}
A.~U. Abeysekara \etal (HAWC Coll.) {\em Astropart. Phys.\/} {\bf 50-52}, 26--32 (2013). (\textit{Preprint} \eprint{1306.5800})
\bibitem{lhaaso1}
G. Di~Sciascio (LHAASO Coll.) {\em Nucl. Part. Phys. Proc.\/} {\bf 279-281}, 166--173 (2016). (\textit{Preprint} \eprint{1602.07600})
   \bibitem{cao2021nat}
Z. Cao \etal (LHAASO Coll.), {\em Nature} {\bf 594}, 33--36 (2021).
\bibitem{stacex-icrc2021} 
G. Rodriguez-Fernandez et al.,  Proc. of 37th International Cosmic Ray Conference (ICRC 2021) (\textit{Preprint} \eprint{2109.08594})
\bibitem{swgo-2019}
Albert A. et al. (SWGO Coll.) (\textit{Preprint} \eprint{arXiv:1902.08429v1})

  \end{thebibliography}
\end{document}